\documentclass[epj]{svjour}

\usepackage{graphicx}

\newcommand{\slL}{\raise.15ex\hbox{$/$}\kern-.53em\hbox{$L$}}
\newcommand{\slP}{\raise.15ex\hbox{$/$}\kern-.53em\hbox{$P$}}
\newcommand{\slp}{\raise.1ex\hbox{$/$}\kern-.63em\hbox{$p$}}
\newcommand{\slk}{\raise.1ex\hbox{$/$}\kern-.63em\hbox{$k$}}
\newcommand{\slq}{\raise.1ex\hbox{$/$}\kern-.63em\hbox{$q$}}
\newcommand{\slv}{\raise.1ex\hbox{$/$}\kern-.63em\hbox{$v$}}
\newcommand{\slR}{\raise.15ex\hbox{$/$}\kern-.53em\hbox{$R$}}
\newcommand{\slQ}{\raise.15ex\hbox{$/$}\kern-.53em\hbox{$Q$}}
\newcommand{\slK}{\raise.15ex\hbox{$/$}\kern-.53em\hbox{$K$}}
\newcommand{\slSigma}{\raise.15ex\hbox{$/$}\kern-.53em\hbox{$\Sigma$}}
\newcommand{\slcalP}{\raise.15ex\hbox{$/$}\kern-.63em\hbox{$\cal P$}}
\newcommand{\slAh}{\hat{\raise.15ex\hbox{$/$}\kern-.73em\hbox{$A$}}}
\newcommand{\slbfA}{\raise.15ex\hbox{$/$}\kern-.73em\hbox{${\imb A}$}}
\newcommand{\slpartial}{\raise.15ex\hbox{$/$}\kern-.53em\hbox{$\partial$}}

\begin{document}

\title{Electron-positron pair production in the external
electromagnetic field of colliding relativistic
heavy ions}
\author{A. Aste\inst{1}, G. Baur\inst{2}, K. Hencken\inst{1},
D. Trautmann\inst{1}
\and G. Scharf\inst{3}
\thanks{Work supported by Swiss National Science Foundation.}
}                     % Do not remove
\institute{Institute for Theoretical Physics, University of Basel,
Klingelbergstrasse 82, 4054 Basel, Switzerland \and
Institut f\"ur Kernphysik (Theorie), Forschungszentrum J\"ulich,
52425 J\"ulich, Germany
\and Institute for Theoretical Physics, University of Z\"urich,
Winterthurerstrasse 190, 8057 Z\"urich, Switzerland}

\date{November 30, 2001}
\abstract{
The results concerning the $e^+e^-$ production in peripheral
highly relativistic
heavy-ion collisions presented in a recent paper by Baltz {\em{et al.}}
are rederived in a very straightforward manner. It is shown that the
solution of the Dirac equation directly leads to the multiplicity, i.e.
to the total number of electron-positron pairs produced by the
electromagnetic field of the ions, whereas the calculation of the
single pair production probability is much more involved.
A critical observation concerns the unsolved problem of seemingly
absent Coulomb corrections (Bethe-Maximon corrections) 
in pair production cross sections.
It is shown that neither the inclusion of the vacuum-vacuum amplitude
nor the correct interpretation of the solution of the Dirac equation 
concerning the pair
multiplicity is able the explain (from a fundamental point of view) the
absence of Coulomb corrections. Therefore the contradiction has to be accounted
to the treatment of the high energy limit.
} %end of abstract
\authorrunning{A. Aste et al.}
\titlerunning{Pair production in external fields}
\maketitle

\section{Introduction}
The problem of $e^+e^-$ pair production by the collision of highly
relativistic nuclei has attracted a lot of interest during the past few years,
since this process will have implications for future experiments
performed at new facilities as BNL's RHIC and CERN's LHC. From the
theoretical side, the result that the Dirac equation can be solved exactly
in the electromagnetic background field created by two nuclei in the limit
where the two nuclei are 'ultrarelativistic' \cite{Segev1,BaltzM},
seemed to lead to the
unexpected consequence that the single pair cross section is equal to its
Born value (given by the second order diagram in Fig. \ref{fig2})
\footnote{
Note that this is only true for the total cross section. As it was shown in 
\cite{HenckenTB}, this is not true for impact parameter dependent
probabilities.}.
The term 'ultrarelativistic' is used here for the limiting
case where the Lorentz factor $\gamma$ becomes large 
($\gamma\rightarrow\infty$), i.~e.
the velocity of the colliding ions approaches the speed of light.
Of course the problem of pair production in an external field goes back to
the beginning of QED \cite{Feynman,Schwinger}. In connection with the
relativistich heavy ion colliders it was found that the impact parameter
dependent probability in perturbation theory can become
larger than one,
which was shown to result in multiple pair production 
\cite{GBaur,HenckenTB,RhoadesBrownW,BestGS,HenckenTB2}.

A series of papers on the same topic followed
\cite{Ivanov,IvanovSerbo,Eichmann,LeeMilstein}
with the aim to show that there must be an error in the
interpretation of the results found in \cite{Segev1}.
An interesting observation was finally made in a paper by Baltz {\em{et al.}}
\cite{Baltz}, where it was shown that the expression for pair production
derived from the results in \cite{Segev1} describes the total number of
produced pairs, and not the single pair production. Additionally, the
importance of taking into account the vacuum-vacuum transition amplitude
when going over from wave mechanics to the full external field
problem with quantized electron field was pointed out, as was already discussed
in \cite{BestGS,HenckenTB2}.
A recent work which treats the structure of Coulomb and unitarity
corrections to the single and multiple pair production is
\cite{LeeMilSer}.

We will derive in this paper
the main results obtained in \cite{Baltz} in a very compact way
using a different field theoretical point of view.
The correct expression for the multiplicity can be derived indeed
in a very straightforward manner from the fundamental equations which
define the S-matrix for the external field problem.
We establish the connection between the single-particle matrix elements
from the solution of the Dirac equation and the pair production in a full
many-body theory. Furthermore, we give a strong argument
which highlights that the results provided from the theory of fermions
in a classical electromagnetic background field
are insufficient to explain the dicrepancies concerning Coulomb corrections 
in the literature.

It is therefore mandatory to understand the correct
interpretation of the results obtained from 'infinite $\gamma$'
calculations and their implications on real electromagnetic processes,
as it was suggested in the work of Lee and Milstein \cite{LeeMilstein}
where they have shown
that Coulomb corrections can be obtained by the correct regularization
of matrix elements.

\section{The external field approximation for QED}
For highly relativistic heavy-ion collisions one usually assumes
that the ions of electric charge $Z_1 e$ and $Z_2 e$
are sufficiently energetic and massive so that the deviation
from straight-line trajectories can be neglected.
The trajectories of the two ions are then, respectively,
$z=\pm t$ ($c=1$), $\vec{x}_\bot = \vec{b}/2$, and there is a gauge in which
the potential of the two ions has the following form \cite{Baltzalone}
\begin{displaymath}
A^\mu(t,\vec{x})=Z_1 e \delta(v_+x)v_+^\mu \log \Bigl(
\frac{(\vec{x}_\bot-\vec{b}/2)^2}{\vec{b}^2} \Bigr)
\end{displaymath}
\begin{equation}
+Z_2 e \delta(v_-x)v_-^\mu \log \Bigl(
\frac{(\vec{x}_\bot+\vec{b}/2)^2}{\vec{b}^2} \Bigr) \quad , \label{efield}
\end{equation}
where $v_\pm^\mu = (1,0,0,\pm 1)/\sqrt{2}$.
The fact that the two electromagnetic fields are compressed due to the
Lorentz contraction to a sheet in the $b$ plane, together with their 
structure in space-time allows for the exact solution of the Dirac equation
in this case.

We will deal in the following with the general external field problem
without referring to the special form of the electromagnetic field
created by ultrarelativistic ions as given by Eq.~(\ref{efield}),
in order to clarify how the solution of the Dirac equation for the electron
wave function and the full S-matrix for the quantized
electron field in an external electromagnetic field are related to
each other. 

\subsection{Construction of the S-matrix}
In order to construct the full S-matrix of QED in the external field
approximation, we start from a one-particle dynamics, defined by a
time-dependent Hamiltonian acting on a state $\Psi$ described by a Dirac
spinor $\Psi(t,\vec{x})$
\begin{equation}
H(t)=H_0+V(t) ,
\end{equation}
where $V(t)$ contains the interaction with the external electromagnetic 
field given by
\begin{equation}
V(t,\vec{x}) \, = \, e (\phi(t,\vec{x})-\vec{\alpha} 
\vec{A}(t,\vec{x})).
\end{equation}
The potentials are assumed to vanish for $t \rightarrow \pm \infty$ in
such a way that the wave operators exist as strong limits
\begin{equation}
W_{{in} \atop {out}} = \lim_{t \rightarrow \pm \infty}
U(t,0)^+ e^{-iH_0t}
\end{equation}
and define a unitary S-matrix $S=W_{out}^+ W_{in}$.
Since we presuppose free dynamics for $t \rightarrow \pm \infty$, we can
base second quantization on the Fock representation of the free Dirac
field.
The one-particle sector of the Fock space then consists naturally
of a positive spectral subspace (electrons) and negative spectral subspace
(positrons) of the Hilbert space of Dirac four-spinors
\begin{equation}
{\cal{H}}_1=
(L^2({\cal{R}}^3))^4=P_+^0{\cal{H}}_1 \oplus P_-^0{\cal{H}}_1 .
\end{equation}
with the standard inner product.
Here, $P_{\pm}^0$ denote the corresponding spectral
projection operators of the free Dirac Hamiltonian.
The quantized Dirac field is then defined as usual in the
Schr\"odinger picture
\begin{equation}
\psi(f)=b(P_+^0f)+d(P_-^0f)^+=b(f_+)+d(f_-)^+ .
\end{equation}
The electron annihilation operator $b(f_+)$
destroys an electron with wave function
$f_+$, and the positron emission operator $d(f_-)^+$
produces a positron with
wave function $f_-$.
The annihilation and emission operators
fulfil the fermionic anticommutation relations
\begin{equation}
\{ b(f_+),b(g_+)^+\}=(f_+,g_+) , \quad \{d(f_-)^+,d(g_-)\}=(f_-,g_-),
\label{commutationrules}
\end{equation}
and the Fock vacuum $\Omega$ is defined by
\begin{equation}
b(f_+) \Omega=d(f_-) \Omega=0 \quad \quad \forall f \in {\cal{H}}_1 \,.
\label{vacuum}
\end{equation}
Note that $b(f_+)$ depends antilinearly on $f_+$, whereas $d(f_-)$
depends linearly on $f_-$. This explains the slight
difference in the commutation relations for electron and positron
operators in (\ref{commutationrules}). 
We mention that naturally, the widely used
field operators can be defined which correspond to the operators
introduced above according to
\begin{equation}
b(f_+)^+=\int d^3x \, b(\vec{x})^+f_+(\vec{x}),
\end{equation}
\begin{equation}
d(f_-)^+=\int d^3x \, d(\vec{x})^+f_-^*(\vec{x})
\end{equation}
or, given an arbitrary orthonormal basis $f_j$ in $P_+^0{\cal{H}}_1$
and $g_k$ in $P_-^0 {\cal{H}}_1$,
\begin{equation}
b^+(\vec{x})=\sum \limits_{j} b(f_j)^+f_j^*(\vec{x}) , \quad
d^+(\vec{x})=\sum \limits_{k} d(g_k)^+g_k(\vec{x})
\end{equation}
such that $\psi(\vec{x})=b(\vec{x})+d(\vec{x})^+$ fulfils the
well-known distributional identity
\begin{equation}
\{ \psi(\vec{x}),\psi(\vec{x}')^+ \}=
{\bf{1}}_4 \delta^{(3)}(\vec{x}-\vec{x}') .
\end{equation}

The one-particle S-matrix, obtained via its perturbation series
\begin{equation}
S=\sum \limits_{n=0}^{\infty} (-i)^n \int \limits_{-\infty}^{+\infty} dt_1
\int \limits_{-\infty}^{t_1} dt_2 \, ... 
\int \limits_{-\infty}^{t_{n-1}} dt_n \, \tilde{V}(t_1)...
\tilde{V}(t_n)
\end{equation}
where
\begin{equation}
\tilde{V}(t)=e^{iH_0t}V(t)e^{-iH_0t} ,
\end{equation}
or by solving directly the Dirac equation with the external field,
can be written as
\begin{equation}
S=\left( \begin{array}{cc}
S_{++} & S_{+-} \\
S_{-+} & S_{--} \\
\end{array} \right) \, , \quad 
S^{-1}=S^+=\left( \begin{array}{cc}
S_{++}^+ & S_{+-}^+ \\
S_{-+}^+ & S_{--}^+ \\
\end{array} \right) \, , \label{defS1}
\end{equation}
if we define $S_{\pm\pm}=P^0_\pm S P^0_{\pm}$ and
$S_{\pm\pm}^+=P^0_\pm S^+ P^0_{\pm}$.
It is often said that $S_{++}$ and $S_{--}$ are related to electron and
positron scattering, whereas $S_{+-}$ and $S_{-+}$ are related to pair
creation and annihilation. But the connection of $S_{\pm \pm}$ to
different measurable quantities has to be treated carefully.
It should be stressed that the one-particle Dirac
theory has no consistent physical interpretation, but only the
corresponding second quantized many-particle theory in Fock space
which we are now going to discuss.

If one lifts the one-particle Hamiltonian
with indefinite energy spectrum to the Fock space
\begin{equation}
{\sf{H}}=\sum \limits_{jk} \Bigl[ (f_j,Hf_k)b(f_j)^+b(f_k)
-(g_j,Hg_k)d(g_k)^+d(g_j) \Bigl] , \label{FockHamiltonian}
\end{equation}
then the Dirac field operator fulfils the Heisenberg equations of motion
\begin{equation}
i\frac{d}{dt} \psi(U^{-1}(t,0)f)=[\psi(U^{-1}(t,0)f), {\sf{H}}],
\end{equation}
\begin{equation}
i\frac{d}{dt} \psi(U^{-1}(t,0)f)^+=[\psi(U^{-1}(t,0)f)^+, {\sf{H}}] .
\label{Hei}
\end{equation}
Roughly speaking, (\ref{FockHamiltonian}) simply accounts for the fact that
the total energy of a general Fock state should be given by the
sum of energies of each single particle. Note also the important
minus sign in (\ref{FockHamiltonian}).

\begin{figure}[htb]
     \centering
     \includegraphics[width=3cm]{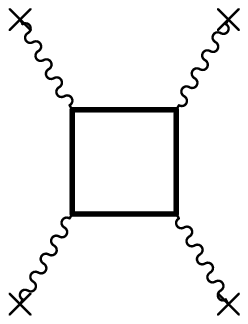}
     \includegraphics[width=10cm]{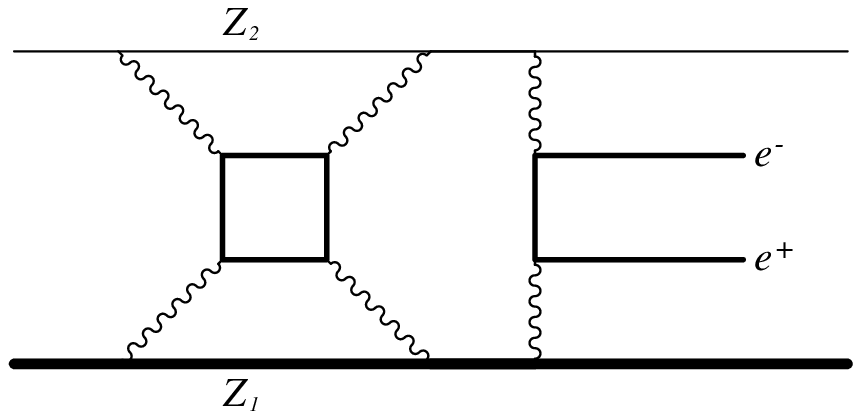}
\caption{Lowest order contribution to the vacuum-vacuum transition amplitude
and single pair production graph containing a vacuum bubble.}
\label{fig1}
\end{figure}

\begin{figure}[htb]
     \centering
     \includegraphics[width=6cm]{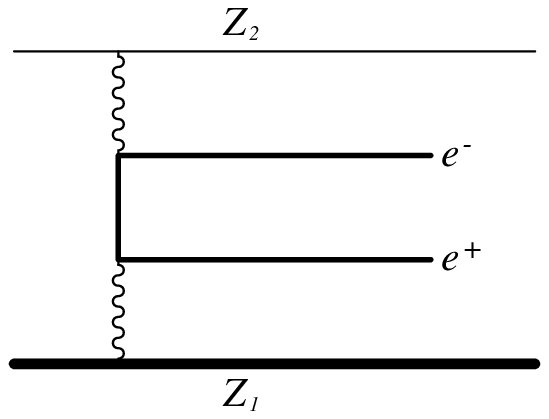}
     \includegraphics[width=6cm]{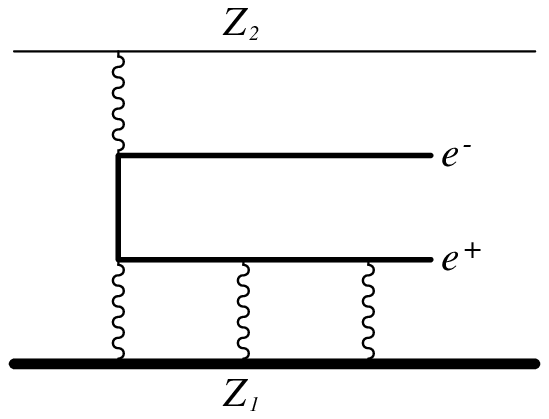}
\caption{Examples of graphs contributing to pair production
for $Z_2 \alpha \ll 1$.}
\label{fig2}
\end{figure}

The definition (\ref{vacuum}) of the Fock vacuum would lead
in the case of time-dependent external fields to Fock representations which
continuously change in time. To avoid this somewhat odd construction,
we retreat to scattering theory by defining
the second quantized S-matrix ${\sf{S}}$ in correspondence to
the Heisenberg equations of motion (\ref{Hei}) by
\begin{equation}
\psi(S^+f)={\sf{S}}^{-1} \psi(f) {\sf{S}} , \label{def1}
\end{equation}
\begin{equation}
\psi(S^+f)^+={\sf{S}}^{-1} \psi(f)^+ {\sf{S}} \label{def2}
\quad \quad \forall f \in {\cal{H}}_1 .
\end{equation}
It is even possible to give an explicit normally ordered
expression for ${\sf{S}}$
\cite{Seipp,Scharf}
\begin{displaymath}
{\sf{S}}=C \, e^{S_{+-}S_{--}^{-1}b^+d^+} : e^{(S_{++}^{+-1}-1)b^+b}:
\times
\end{displaymath}
\begin{equation}
:e^{(1-S_{--}^{-1})dd^+}:e^{S_{--}^{-1}S_{-+}db} ,
\label{eq_Smatrix}
\end{equation}
where the terms in the exponents are a shorthand for the operators
\begin{equation}
Ab^+b \equiv \sum \limits_{jk} (f_j,Af_k)b(f_j)^+ b(f_k) ,
\label{lift}
\end{equation}
\begin{equation}
Ab^+d^+ \equiv \sum \limits_{jk} (f_j,Ag_k)b(f_j)^+ d(g_k)
\end{equation}
and similarly for $Add^+$ and $Adb$.
Note that $S_{--}^{-1}$ is the inverse of the matrix $S_{--}$ alone
and therefore not identical to $S_{--}^+=P^0_{-} S^{-1} P^0_{-}$.
$C$ denotes the vacuum-vacuum transition amplitude
$(\Omega,{\sf{S}} \Omega)$. 
This vacuum-vacuum transition amplitude is always present, but whereas it
has absolute value one, e.g., for static fields (and is therefore dropped
in the calculations), $C$ is indispensable for the unitarity
of ${\sf{S}}$, as has been verified in \cite{Baltz} by perturbation
theory. In our case where two static Coulomb fields
are moving relatively to each other, the lowest order contribution
to the vacuum-vacuum transition amplitude is given by a loop graph shown
in Fig.\ref{fig1}. It is a well-known result that all vacuum graphs
appearing in pair production amplitudes (see the example in Fig. \ref{fig1})
can be absorbed in the factor $C$. Additionally, the lowest order contribution
is finite due to gauge invariance, whereas higher order contributions
to the vacuum-vacuum transition amplitude are finite due to power
counting arguments. Therefore, the theory is free of any regularization
problems.

Equation~(\ref{eq_Smatrix}) is the important link showing the relation between
the (single-particle) $S$ matrix from the solution of the Dirac equation and
the general many-body theory. 

We finally mention that the S-matrix ${\sf{S}}$ in Fock space exists,
if and only if $S_{+-}$ is a Hilbert-Schmidt operator. This condition
is not satisfied with the potential (\ref{efield}) as it stands, but
it can be achieved by appropriate regularization.

\subsection{Multiplicity}
From (\ref{def1}) it follows immediately
\begin{equation}
b(f){\sf{S}}={\sf{S}}[b(S_{++}^+f)+d(S_{-+}^+f)^+] , \label{comm1}
\end{equation}
a simple identity which allows to derive easily the
average number of electron-positron pairs produced by the external field.
The number of pairs in the final state is equal to the number of electrons
in the final state, which can be calculated from the electron 
(or positron) number operator
\begin{equation}
N_e=\sum \limits_{n} b(f_n)^+b(f_n) , \quad
N_p=\sum \limits_{n} d(g_n)^+d(g_n) .
\end{equation}
Therefore, we obtain for the number of pairs in the final state
created by the external field out of the vacuum
${\sf{S}} \Omega$
\begin{displaymath}
N=({\sf{S}} \Omega , \sum \limits_{n} b(f_n)^+ b(f_n) {\sf{S}} \Omega)
\end{displaymath}
\begin{equation}
=\sum \limits_{n}( b(f_n) {\sf{S}} \Omega , b(f_n) {\sf{S}} \Omega) .
\end{equation}
Using (\ref{comm1}), this can be written as
\begin{displaymath}
N=\sum \limits_{n} (d(S_{-+}^+f_n) \Omega , d(S_{-+}^+f_n) \Omega)=
\sum \limits_{n} (f_n, S_{+-}S_{-+}^+ f_n)
\end{displaymath}
\begin{equation}
=\mbox{Tr}(S_{+-}S_{-+}^+)
\label{multiplicity} \end{equation}
or equivalently, using the positron number operator
\begin{equation}
N=\mbox{Tr}(S_{-+}S_{+-}^+) . \label{posinum}
\end{equation}
Therefore, the results obtained from the solution of the Dirac equation
for $S_{-+}$ \cite{Segev1}
must not be misinterpreted as the single pair production
amplitude. Furthermore, this result for which a lengthy
derivation has been given in \cite{Baltz}, follows here in an
almost trivial way.

\subsection{Single pair production probability}
In order to derive the production probability for a single
electron-positron pair with wave functions $\Phi_e$ and $\Phi_p$,
we have to compute first the S-matrix element
\begin{displaymath}
M_{ep}=(b(\Phi_e)^+d(\Phi_p)^+ \Omega, {\sf{S}} \Omega)
\end{displaymath}
\begin{equation}
=C(\Omega, d(\Phi_p)^+ b(\Phi_e)^+ e^{S_{+-}S_{--}^{-1}b^+d^+} \Omega)
\quad .
\end{equation}
The definition (\ref{lift}) and the anticommutation rules
(\ref{commutationrules}) enable us after some algebra to derive
the commutation rules \cite{Scharf}
\begin{displaymath}
b(f)e^{S_{+-}S_{--}^{-1}b^+d^+}
\end{displaymath}
\begin{equation}
=e^{S_{+-}S_{--}^{-1}b^+d^+}
(b(f)+d((S_{+-}S_{--}^{-1})^+  P_+^0 f)^+) \quad ,
\end{equation}
\begin{displaymath}
d(f)e^{S_{+-}S_{--}^{-1}b^+d^+}
\end{displaymath}
\begin{equation}
=e^{S_{+-}S_{--}^{-1}b^+d^+}
(d(f)-b(S_{+-}S_{--}^{-1}  P_-^0 f)^+) \quad ,
\end{equation}
which reduce the matrix element to the form
\begin{displaymath}
M_{ep}=C(\Omega, d(\Phi_p) d((S_{+-}S_{--}^{-1})^+ P_+^0 \Phi_e)^+ 
\Omega)
\end{displaymath}
\begin{equation}
=C(\Phi_e,S_{+-}S_{--}^{-1} \Phi_p) .
\end{equation}
The matrix ${\cal{M}}=S_{+-}S_{--}^{-1}$
contains the additional scattering factor
$S_{--}^{-1}$, which is missing in the literature when differential
cross sections for pair production are considered.
As will be explained below, ${\cal{M}}$ is identical to the result using
the usual Feynman propagators.

Unfortunately, the calculation of an integral operator like
$S_{--}^{-1}$ is not a trivial task, but from Eq.~(\ref{defS1})
we can derive the unitarity condition
\begin{equation}
S_{-+} S_{+-}^+ + S_{--} S_{--}^+ = P_-^0 \, ,
\end{equation}
which can be multiplied from the left by $S_{+-} S_{--}^{-1}$
leading to the equation
\begin{equation}
{\cal{M}}=S_{+-} S_{--}^+ + {\cal{M}} S_{-+} S_{+-}^+ \, , \label{recur}
\end{equation}
which can be iterated.

\subsection{n-pair production}
The vacuum-vacuum transition probability $|C|^2$ can be derived
from
\begin{equation}
1=({\sf{S}}\Omega,{\sf{S}}\Omega)=|C|^2 (e^{{\cal{M}}b^+d^+} \Omega,
e^{{\cal{M}}b^+d^+} \Omega) ,
\end{equation}
with the result \cite{Seipp}
\begin{equation}
|C|^{-2}=\mbox{det}(1+{\cal{M}}^+{\cal{M}}) .
\end{equation}
It is clear that the probability $P_{tot}$ to produce an arbitrary
number of pairs is
equal to $1-|C|^2$. The well-known identity
\begin{equation}
\log \mbox{det} (1+{\cal{M}}^+{\cal{M}})=\mbox{Tr}
\log (1+{\cal{M}}^+{\cal{M}})
\end{equation}
leads to
\begin{equation}
\frac{1}{|C|^2}=\exp\Biggl(
-\mbox{Tr} \sum \limits_{n=1}^{\infty}
\frac{(-{\cal{M}}^+{\cal{M}})^n}{n} \Biggr) \, ,
\end{equation}
or
\begin{equation}
\frac{1}{|C|^2} \sum \limits_{n=1}^\infty P_n = \frac{1}{|C|^2}-1=
\exp \Biggl(-\mbox{Tr} \sum \limits_{n=1}^{\infty}
\frac{(-{\cal{M}}^+{\cal{M}})^n}{n} \Biggr)-1 . \label{summed}
\end{equation}
Utilizing the expansion in (\ref{summed}) allows to write down
the n-pair production probability
 by collecting all terms
in (\ref{summed}) containing n ${\cal{M}}^+{\cal{M}}$
terms, e.g.
\begin{equation}
P_2=\frac{|C|^2}{2}[(\mbox{Tr}({\cal{M}}^+{\cal{M}}))^2-\mbox{Tr}
(({\cal{M}}^+{\cal{M}})^2)] . \label{twopairs}
\end{equation}
This can be understood if one takes a look at the actual
calculation of the n-pair production probability analogous to
the single pair production calculation presented in sect. 2.3.
It follows also from the expression of the determinant in terms
of antisymmetric tensor products which can be found in \cite{Reed}
\begin{equation}
\mbox{det}(1+A)=\sum \limits_{n=0}^{\infty} \mbox{Tr}(\Lambda^n(A)) ,
\end{equation}
if $A$ is a trace-class operator. $\Lambda^n(A)$ is the tensor product
of $A$ over the antisymmetric n-particle space.

If 'ex\-change terms' of the form $\mbox{Tr}(({\cal{M}}^+{\cal{M}})^n)$
are neglected in (\ref{summed}),
then the probability of producing n pairs goes over into
a Poisson distribution
\begin{equation}
P_n \rightarrow e^{-{\mbox{Tr}}({\cal{M}}^+{\cal{M}})}
\frac{(\mbox{Tr}({\cal{M}}^+{\cal{M}}))^n}{n !},
\end{equation}
which seems to be a reasonable approximation
due to the correlation between the momenta of electron and positron in this
case, as was also found in the calculations
performed in \cite{HenckenPoisson}. The poisson distribution had
already been derived theoretically in an earlier work \cite{GBaur}.
There, the sudden (or Glauber) approximation and a quasiboson
approximation for $e^+e^-$ pairs was assumed, leading to a simple
but very instructive example which elucidates the pair production
mechanism. It is remarkable that the Poisson distribution gives the
multiplicity of particles to be
\begin{equation}
N_{Poisson}={\mbox{Tr}}({\cal{M}}^+{\cal{M}})
\end{equation}
quite similar to the exact one of Eq.~(\ref{posinum}).

\subsection{Perturbation theory}
We briefly mention here the perturbative expansion of the previously
discussed matrix elements.
As it follows from the solution of the Dirac equation
the perturbative expansion of $S_{\pm \pm}$ is given by retarded
propagators
\begin{equation}
S^{ret}(p)=\frac{\slp+m}{p^2-m^2+ip_00} ,
\end{equation}
\begin{displaymath}
S_{\pm \pm}^{(n)}(\vec{p},\vec{q})=
-i\frac{e^n}{(2 \pi)^{2n-1}}P_{\pm}^0(\vec{p}) \gamma^0
\int d^4p_1 \, ... \, d^4p_{n-1}
\end{displaymath}
\begin{equation}
\slAh (p-p_1)S^{ret}(p_1) \slAh(p_1-p_2)
\, ... \, S^{ret}(p_{n-1}) \slAh(p_{n-1}-q) P_{\pm}^0(\vec{q}) .
\label{Retardedpert}
\end{equation}
The projection operators are given in momentum space by
\begin{equation}
P_\pm^0(\vec{p})=\frac{1}{2E}(E \pm \vec{\alpha}\vec{p}+\beta m) ,
\end{equation}
where $\vec{\alpha}$ and $\beta$ are standard Dirac matrices.
The pair production amplitude, on the contrary, has to be calculated
using Feynman propagators
\begin{equation}
S^F(p)=\frac{\slp+m}{p^2-m^2+i0} ,
\end{equation}
\begin{displaymath}
(S_{+-}S_{--}^{-1})^{(n)}(\vec{p},\vec{q})
\end{displaymath}
\begin{displaymath}
=-i\frac{e^n}{(2 \pi)^{2n-1}}P^0_{+}(\vec{p}) \gamma^0
\int d^4p_1 \, ... \, d^4p_{n-1}
\end{displaymath}
\begin{equation}
\slAh (p-p_1)S^{F}(p_1) \slAh(p_1-p_2)
\, ... \, S^{F}(p_{n-1}) \slAh(p_{n-1}-q) P^0_{-}(\vec{q}) .
\label{Feynmanpert}
\end{equation}
Eqs.~(\ref{Retardedpert},\ref{Feynmanpert})
represent well-known results from
(quantum) field theory which are discussed in many standard textbooks
\cite{Itzykson,Bogoliubov}.
An explicit calculation which shows how the transition from retarded to
Feynman propagators takes place when $S_{\pm \pm}^{(n)}$ and
$(S_{+-}S_{--}^{-1})^{(n)}$ are compared can be found in \cite{Scharf}.
It is the inverse of $S_{--}$ which is responsible for this transition.

Since it is possible to solve the Dirac equation exactly in the
ultrarelativistic limit, it is clear that the retarded
propagator is also known in this case, and is therefore a much
simpler object than the Feynman propagator from the calculational
point of view.

\section{Coulomb corrections and multiplicity}
It was argued in \cite{Baltz} that the absence of Coulomb corrections
in the high energy limit as derived in \cite{Segev1}
could be attributed to the
difference in the interpretation between the multiplicity 
$N={\mbox{Tr}}(S_{+-}S_{-+}^+)$ and the single pair production
probability
$P_1=|C|^2 {\mbox{Tr}}({\cal{M}}^+ {\cal{M}})$.
Here we want to present some arguments
that show that such an explanation cannot hold. Coulomb corrections and
'unitarity corrections' (following the terminology used in
\cite{LeeMilSer}) can be treated as two different 
phenomena,
which stem from different sources. Whereas unitarity corrections are due
to many-particle aspects, Coulomb corrections affect the matrix element
${\cal{M}}$ directly.

Since the theory of pair production developed above is valid quite generally,
we can consider the special case
with two ions that have charge numbers $Z_1$ and $Z_2$, in the limit where
$Z_1 \alpha$ remains fixed but $Z_2 \alpha$ goes to zero.
Expanding in powers of $Z_2\alpha$ and realizing
that the potential of the ion with charge $Z_1 e$
is a static Coulomb potential in
its rest frame, we find that both $S_{-+}$ and $S_{+-}$ are at least of order
$Z_2\alpha$. Using the recursion relation of Eq.~(\ref{recur}),
the second term on
the right hand side is at least of order $(Z_2 \alpha)^3$ and can therefore be
neglected. Therefore, if only contributions at first order in $Z_2 \alpha$
are kept (whereas contributions at higher orders in $Z_1 \alpha$ are
important, as it is illustrated in Fig. 2),
the matrix element for pair production is found to be
${\cal{M}}\approx S_{+-}S_{--}^+$. This can also be seen directly from 
the definition ${\cal{M}}=S_{+-}S_{--}^{-1}$ and the fact that for a static
field we have no off-diagonal elements in $S$ with respect to the block
structure given in Eq. (\ref{defS1}) and therefore 
$S_{--}^{-1}=S_{--}^+$. Summing over all final states we get for $P_1$
(keeping in mind that at this order of $Z_2\alpha$ also $|C|^2=1$)
\begin{equation}
P_1 \approx {\mbox{Tr}}(S_{+-}S_{--}^+S_{--}S_{-+}^+)
= {\mbox{Tr}}(S_{+-}S_{-+}^+).
\end{equation}
Following the arguments for the high energy limit of \cite{Segev1} this
probability should now be identical to the lowest order result. On the
other hand the limit performed here corresponds exactly to the one
which is done in order
to derive the Bethe-Maximon theory \cite{BetheMaximon,Ivanov}. Therefore the 
discrepancy between the two results is also present in this case,
independently
of the many-particle aspects. Comparing the two multiplicities $N$ and 
$N_{Poisson}$ one finds that they start to disagree only at fourth order in
$Z_2 \alpha$.

Therefore one has to conclude that the discrepancies in
the literature concerning
Coulomb corrections cannot be traced
back to the many-body aspects of pair production in this case.

\section{Conclusion and Outlook}
In this paper we have derived in a very straightforward way the link between
the single-particle $S$ matrix elements
obtained from the solution of the Dirac equation
and its relation to the multiplicity and single-pair production probability.
The term $S_{--}^{-1}$, which is missing in many papers
and which is important for the calculation of the differential
cross section with respect to both electrons and positrons, was
introduced.
A careful analysis of the limit which
is relevant for a comparison with the Bethe-Maximon results shows that the
absence of Coulomb corrections in this high energy limit cannot be understood
by a purely conceptional treatment of the external field problem.
The problem therefore seems to lie in the way this high energy
limit is made. That a too
simplified use of the eikonal approximation leads to a result in contradiction
with the Bethe-Maximon one was already pointed out in \cite{Blankenbecler},
whereas within a more refined analysis as given in \cite{Kopeliovich} there
is no disagreement.

In priniciple the pair production in external fields is well understood,
but the numerical evaluation of the Cou\-lomb cor\-rec\-tions for two
highly charged ions, especially at small impact parameter,
still needs to be found.


\begin{thebibliography}{}
\bibitem{Segev1}
B. Segev, J.C. Wells, Phys. Rev.~C 
{\textbf{59}}, no. 5, 2753 (1999)
\bibitem{BaltzM}
A.~J. Baltz, L. McLerran, Phys. Rev.~C {\textbf{58}},  1679  (1998)
\bibitem{Feynman}
R.~P. Feynman, Phys. Rev. {\textbf{76}},  749  (1949)
\bibitem{Schwinger}
J. Schwinger, Phys. Rev. {\textbf{93}},  615  (1954)
\bibitem{GBaur}
G. Baur, Phys. Rev.~A {\textbf{42}}, no. 9, 5736 (1990)
\bibitem{HenckenTB}
K. Hencken, D. Trautmann, G. Baur, Phys. Rev.~C {\textbf{59}}, 841 (1999)
\bibitem{RhoadesBrownW}
M.~J. Rhoades-Brown, J. Weneser, Phys. Rev.~A {\textbf{44}},  330  (1991)
\bibitem{BestGS}
C. Best, W. Greiner, G. Soff, Phys. Rev.~A {\bf 46},  261  (1992)
\bibitem{HenckenTB2}
K. Hencken, D. Trautmann, G. Baur, Phys. Rev.~A {\bf 51},  998  (1995)
\bibitem{Ivanov}
D.Yu. Ivanov, K. Melnikov, Phys. Rev.~D {\textbf{57}}, 4025 (1998)
\bibitem{IvanovSerbo}
D.Yu. Ivanov, A. Schiller, V.G. Serbo, Phys. Lett.~B
{\textbf{454}}, 155 (1999)
\bibitem{Eichmann}
U. Eichmann, J. Reinhardt, W. Greiner, Phys. Rev.~A {\textbf{61}},
062710 (2000)
\bibitem{LeeMilstein}
R.N. Lee, A.I. Milstein, Phys. Rev.~A {\textbf{61}}, 032103 (2000)
\bibitem{Baltz}
A.J. Baltz, F. Gelis, L. McLerran, A. Peshier, Nucl. Phys.~A
{\textbf{695}}, 395 (2001)
\bibitem{LeeMilSer}
R.N. Lee, A.I. Milstein, V.G. Serbo, hep-ph/0108014
\bibitem{Baltzalone}
A.J. Baltz, Phys. Rev. Lett. {\textbf{78}}, no. 7, 1231 (1997)
\bibitem{Seipp}
H. P. Seipp, Helvetica Physica Acta {\textbf{55}}, 1 (1982)
\bibitem{Scharf}
G. Scharf, Finite quantum electrodynamics: the causal approach
(Springer, New York, 1995) (second edition)
\bibitem{Reed}
M. Reed, B. Simon, Methods of Modern Mathematical Physics
Vol. IV: Analysis of Operators (Academic Press, New York, 1978),
p. 323, eq. (188)
\bibitem{HenckenPoisson}
K. Hencken, D. Trautmann, G. Baur, Phys. Rev.~A {\textbf{51}},
998 (1995)
\bibitem{Itzykson}
C. Itzykson, J.-B. Zuber, Quantum Field Theory (McGraw-Hill, 1980), p. 89
\bibitem{Bogoliubov}
N.N. Bogoliubov, D.V. Shirkov, Introduction to the Theory of Quantized
Fields (Interscience Publishers, Inc., New York, 1959)
\bibitem{BetheMaximon} H.A. Bethe, L. Maximon,
Phys. Rev. {\textbf{93}}, 768 (1954)
\bibitem{Blankenbecler}
R. Blankenbecler, S.~D. Drell, Phys. Rev.~D {\textbf{36}}, 2846 (1985)
\bibitem{Kopeliovich}
B.Z. Kopeliovich, A.V. Tarasov, O.O. Voskresenskaya, Eur. Phys. J.~A
{\textbf{11}}, 345 (2001)
\end{thebibliography}
\end{document}